\documentclass{elsart}
\usepackage{epsfig}
\newcommand{\beq}{\begin{eqnarray}}
\newcommand{\eeq}{\end{eqnarray}}

\begin{document}

\begin{frontmatter}

\title{Appearence of Mother Universe 
and Singular Vertices in Random Geometries}

\author[Amsterdam]{Piotr Bialas\thanksref{pperm}},
\author[Bielefeld]{Zdzislaw Burda\thanksref{zperm}},
\author[Bielefeld]{Bengt Petersson} \\ and
\author[Bielefeld]{Joachim Tabaczek}

\address[Amsterdam]{
{Universiteit van Amsterdam, Instituut voor Theoretische
Fysica,}\\ 
{ Valckenierstraat 65, 1018 XE Amsterdam, The
Netherlands}} 
\thanks[pperm]{Permanent address: Institute of Comp. Science,
Jagellonian University,\\ ul. Nawojki 11, 30-072 Krak\'ow, Poland}
\address[Bielefeld]{Fakult\"{a}t f\"{u}r Physik, Universit\"{a}t
Bielefeld,\\ Postfach 10 01 31, Bielefeld 33501, Germany}
\thanks[zperm]{Permanent address: Institute of Physics, 
Jagellonian University, ul. Reymonta 4, 30-059 Krak\'{o}w, Poland}

\begin{abstract}

We discuss a general mechanism that drives the phase transition 
in the canonical ensemble in models of random geometries. 
As an example we consider a solvable model of branched polymers
where the transition leading from tree-- to bush--like polymers
relies on the occurrence of vertices with a large number of
branches. The source of this transition is a combination
of the constraint on the total number of branches in the canonical
ensemble and a nonlinear one--vertex action.  We argue that exactly 
the same mechanism, which we call {\em constrained mean--field}, 
plays the crucial role in the phase transition in 4d simplicial gravity
and, when applied to the effective one--vertex action, explains the 
occurrence of both the mother universe and singular vertices at the 
transition point when the system enters the crumpled phase. 
\par \mbox{} \par \noindent
ITFA-96-29, BI-TP 96/32, hep-lat/9608030
\end{abstract}

\end{frontmatter}

\section*{Introduction}

Simplicial quantum gravity in four dimensions is known to have two
geometrically distinct phases called, in a manner reflecting their
basic geometrical features, respectively the elongated and the
crumpled phase \cite{am,aj1}. For large values of the coupling
constant, frequently denoted by $\kappa_2$, that controls the
curvature of the universe and, roughly speaking, corresponds to a
regularized version of the inverse Newton constant, the typical
universe has a tree--like shape. The branches of this tree consist of
so--called ``baby universes'' that are linked to each other by very narrow
bottlenecks resembling the wormholes known from other considerations
of Euclidean Quantum Gravity \cite{c}. In this phase, simplicial
gravity has been shown to behave like an ensemble of branched polymers
with the characteristic values of a Hausdorff dimension $d_H = 2$ and
an entropy exponent $\gamma = 1/2$ \cite{aj2}. For $\kappa_2$ below the
critical value \cite{bbkp1}, simplicial gravity enters the crumpled phase. 
So--called singular vertices occur on the typical geometry, vertices of a large
order that grows linearly with the volume of the ensemble \cite{hin,ckrt}.
Because of this, almost the whole universe is in the neighbourhood
of these singular vertices, which means that the average geodesic
distances between simplices depends weakly if at all on the universe's
volume. It is therefore frequently stated that the crumpled phase has a very
large or even infinite Hausdorff dimension. The appearance of the
singular vertices coincides with a dramatic change in the tree of
baby universes; in the elongated phase all the baby universes have
the same status and none of them is favoured, whereas here in the crumpled
phase one of them becomes a ``mother universe'' which has a much
larger volume than the others and has many baby universes
directly linked to it. As we will show, this effect can be easily
understood in terms of a constrained mean--field used as an effective
model for the tree of baby universes. There is also a clear correlation
between the appearance of singular vertices and the typical baby
universe structure, as was for example shown in \cite{bbkp2} where the
correlations between the maximal order of the vertex and the average
minbu size were observed at the transition point. It is, for instance,
seen that the mother universe always contains the singular vertex.
Because of these correlations, we will argue that the occurrence of
singular vertices is also caused by the mechanism discussed here.

\section*{Branched Polymers and Constrained Mean--field Models}

It has been well established by now that the branched polymer model very
accurately describes the elongated phase of 4d simplicial gravity
\cite{aj2,b,b2}. We will argue that this description in terms of branched
polymers can be extended beyond the phase transition into the crumpled
phase. First, however, let us define this model and its generalizations,
which we call constrained mean--field models.

The branched polymer model describes an ensemble of abstract trees
which link different vertices via branches in such a way that there
are no closed loops on the corresponding graph. To each vertex we
assign a {\em vertex order} or {\em branching number} $n$, which is
defined as the number of branches emerging from this vertex. There is a
certain given probability distribution $p(n)$ for the branching number at
each vertex, but the numbers of branches at any two different vertices
are independent. If, however, one considers the canonical ensemble
of trees with a fixed number of vertices, say $N$, then the total
number of branches on the tree is fixed at $2N-2$ by the Euler relation.
This induces an effective dependence between the vertex orders, 
results in the multi--point probability
\beq
p(n_1,n_2,\dots,n_N) \sim p(n_1)\cdots p(n_N)
\delta_{n_1+\cdots+n_N,2N-2} 
\label{e1}
\eeq
and leads to geometrical correlations on the branched polymers \cite{b,b2}.
This model, for so--called planar rooted trees and with the power--like
probability $p(n) \sim n^{-\beta}$ for the branching number, has been solved
in \cite{bb}. At a certain value of $\beta$ the model was shown
to undergo a phase transition between two phases called, respectively,
the tree-- and bush--like phases. The tree phase, which occurs for
$\beta < \beta_c$, is a genuine branched polymer phase. Above the
critical $\beta$, however, one of the vertices becomes singular and
the system enters the bush--like phase, where most branches grow directly
out of the singular vertex, or root of the bush.
\begin{figure}[t]
\begin{center}
\epsfig{file=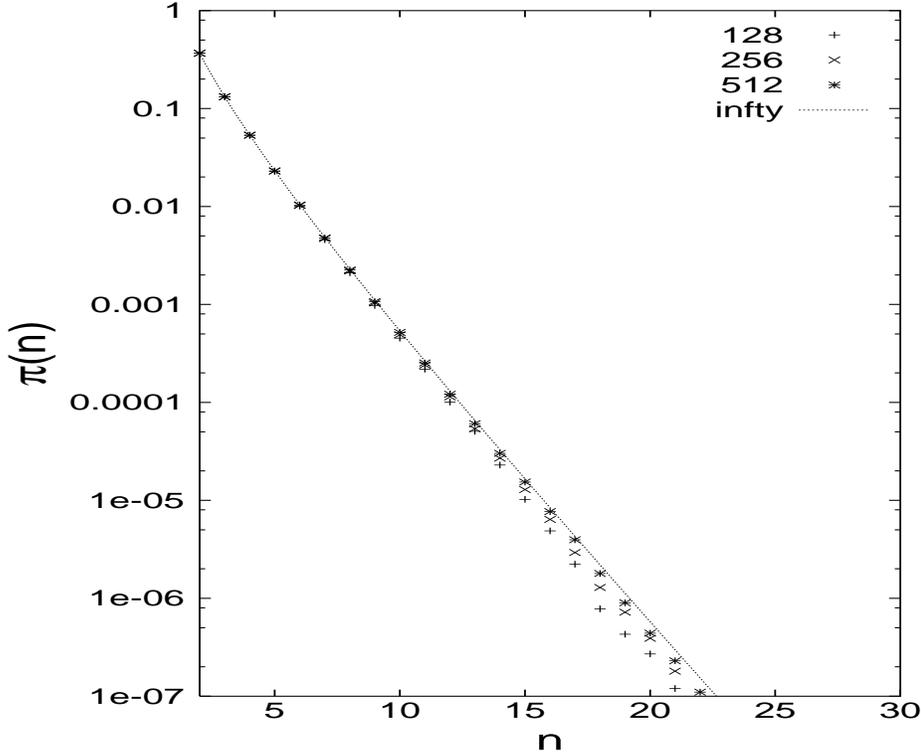,height=10cm,width=12cm}
\end{center}
\caption{\label{f1}Distribution of the branching number in the tree 
phase of BP.}
\end{figure}
\begin{figure}[t]
\begin{center}
\epsfig{file=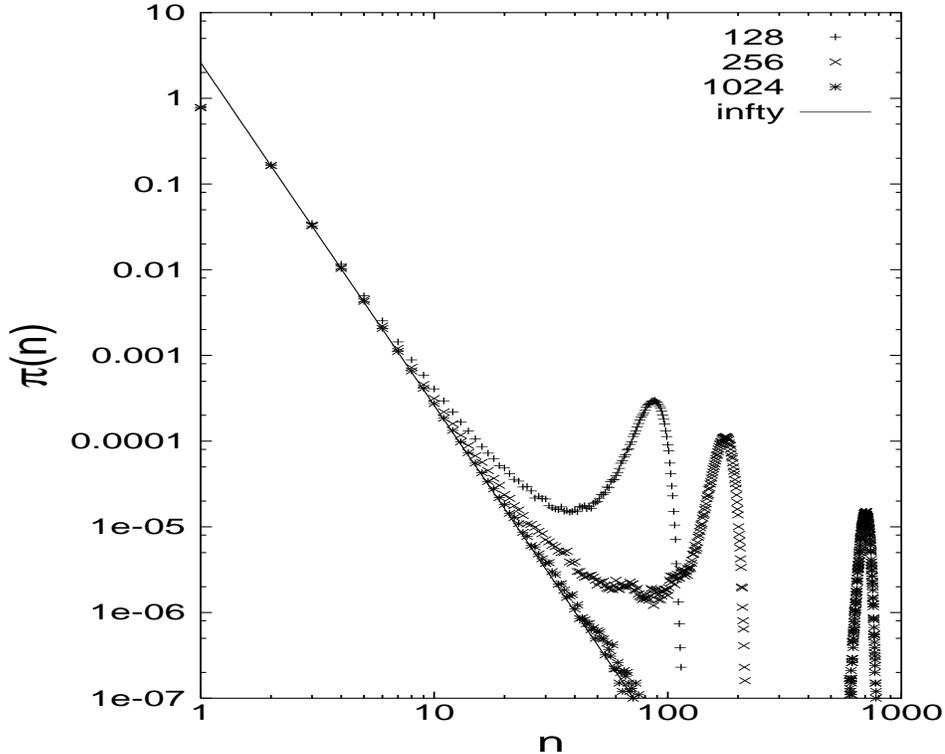,height=10cm,width=12cm}
\end{center}
\caption{\label{f2}Distribution of the branching number in the 
bush phase of BP.}
\end{figure}
In figure \ref{f1} we show the normalized distribution $\pi(n)$ 
of vertex orders for different tree sizes $N$. With increasing $N$ the
distributions approach smoothly the one calculated in the
thermodynamic limit. In figure \ref{f2} we show distributions for
different polymer sizes in the bush phase. One clearly sees the appearance
of a singular vertex, or bush root. Moreover, the order of this vertex grows
linearly with the polymer size; one could say that the bush grows only
around its root. Looking at the peak in the probability distribution
$\pi(n)$ that corresponds to the root, one can also check that it
becomes smaller in inverse proportion to the polymer size, which means
that independently of the size there is always the same number of
roots per tree (in this case one).  The detailed calculations of the
vertex order distribution and the full discussion will be
presented in the forthcoming publication \cite{bb2}.

The behaviour described here is typical of a large class
of models which we call constrained mean--field (or non--interacting
constrained) models. We will not give a strict definition of this
class, but rather present another, very simple model, which captures all
its essential characteristics. The mathematical structure of this model
is the same as the branched polymer's and will neatly explain the
behaviour discussed above.

Suppose that we have a lattice model describing a certain local
quantity whose states can be enumerated by a discrete positive number
$n_i$, where the index $i$ runs over lattice points on which the
quantity is defined. We assume that this model does not contain any
explicit interactions, {\em i.e.} the action is just a simple sum over
all the points of the lattice~: $S(n)=\sum_{i=1}^{N}s(n_i)$.  As it
stands, this model is, of course, trivial; but we put in one additional
constraint~: $\sum_{i=1}^N n_i=M$. One realisation of this model would
be a system of $M$ balls in $N$ boxes where the one--box action
depends only on the number of balls in the box. The partition function
of this model is
\begin{eqnarray} 
Z(N,M)=\sum_{n_1,\ldots,n_N}p(n_1)\cdots p(n_N)
\delta_{n_1+\cdots+n_N,M}
\label{znm}
\end{eqnarray}
where $p(n_i) = e^{-s(n_i)}$.

Without the constraint the one--point effective (or dressed) probability
$\pi(n)$ would just be equal to the bare probability $p(n)$. Because of the
constraint, however, it is given by
\begin{eqnarray}\label{pi}      
\pi(n)=p(n)\frac{Z(N-1,M-n)}{Z(N,M)}
\end{eqnarray}
It is worth noting here that the dressed one--point probability 
$\pi(n)$ will not change if we add to the action a term linear in
$n_i$ ({\em i.e.} $-\mu n_i$), as such a term will cancel itself in the
numerator and denominator in formula (\ref{pi}). This is actually a
very crucial point because it means that it is only the combination
of a constraint on the sum of $n_i$s and an action that is non--linear
in the $n_i$s that makes this model non--trivial.

\begin{figure}[t]
\begin{center}
\epsfig{file=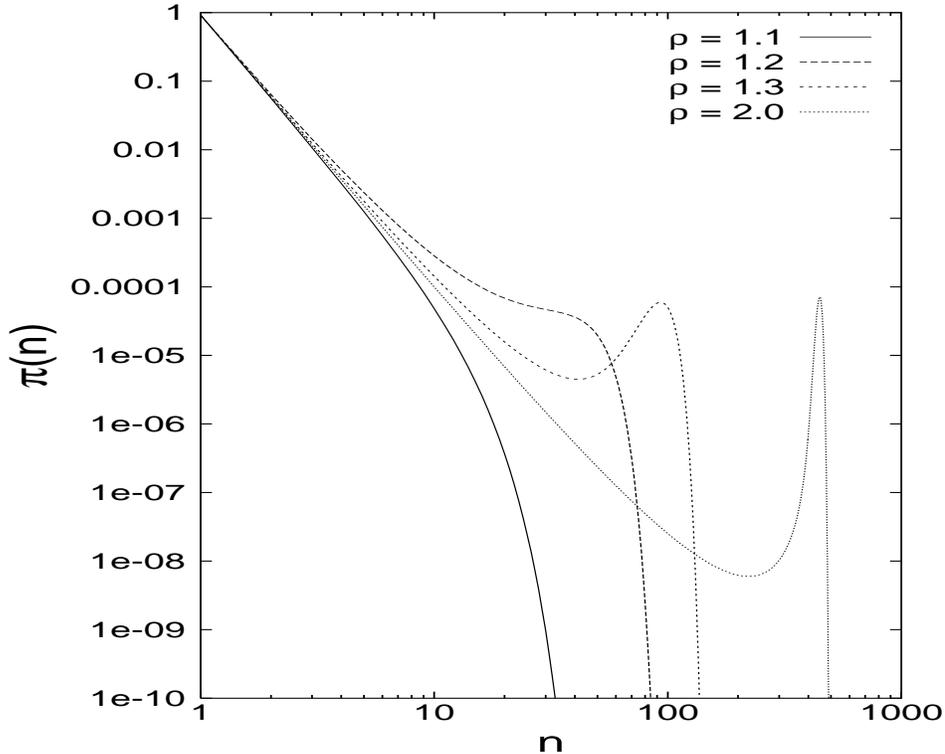,height=10cm,width=12cm}
\end{center}
\caption{\label{f3} The density induced transition 
in the balls--in--boxes model.}
\end{figure}

If we choose $p(n) \sim n^{-\beta}$ and $M=\rho N$, we
find that, similarly to the branched polymer case, the model has
two phases~: a ``distributed/fluid'' phase where every box has, on average,
$M/N$ balls; and a ``condensed'' phase where one box will contain a
number of balls proportional to $N$, resulting in a peak in the
dressed box occupation probability. Obviously, the situation is the same
as above. In fact, the branched polymer model can
be mapped exactly onto this type of balls--in--boxes model \cite{bb2}.
However, the balls--in--boxes model is more general because we can vary the
density $\rho$ of the balls. It turns out that for fixed $\beta$
the transition can also be produced by changing the density of the
balls \cite{bb2}. By increasing the density we can bring the system into
the condensed phase (see fig. \ref{f3}).
 
\section*{The Minbu Tree}

The physical picture which explains the equivalence of the
elongated phase of gravity and branched polymers is that the typical
geometry of simplicial gravity is effectively described by the cascade
of baby universes weakly interacting through the wormholes. With each
elongated configuration of simplicial gravity one can associate an abstract
tree in which each baby universe is represented by a vertex and each
wormhole by a branch joining those two vertices that correspond to the
baby universes which are linked through this wormhole.

The average number $N$ of vertices on this tree depends on the number $N_4$
of simplices on the triangulation. However, only $N_4$ is actually fixed
in canonical simulations of simplicial gravity, and $N$ will fluctuate
around the average value given by $N_4$. So for the tree of baby universes
we do not have a true constraint as we had on the branched polymer,
and the formula \ref{e1} cannot be expected to be more than a good
approximation. We will show that it is indeed very good.

We will, in the following, directly study the underlying
tree structure of baby universes. More specifically, we will restrict
ourselves to the tree of so--called minbus, which are MInimal
Neck Baby Universes \cite{jm}. The smallest 
bottleneck possible on a four--dimensional simplicial manifold
consists of five tetrahedra connected to each other in such a way that they  
form the sceleton of a four--simplex that is not itself a part of the
simplicial manifold. Each minbu can have outgrowths on itself, which are
then considered its neighbouring minbus. If one confines oneself to
spherical topology, one can build a generation tree by pointing from each
minbu to its respective 'parent'. More precisely, one starts from all
minbus that have only one wormhole. These form the last generation and
are linked by their wormholes to their respective fathers, which make up the
next--to--last generation. Repeating this procedure, one recovers the whole
tree.
\begin{figure}[t]
\begin{center}
\epsfig{file=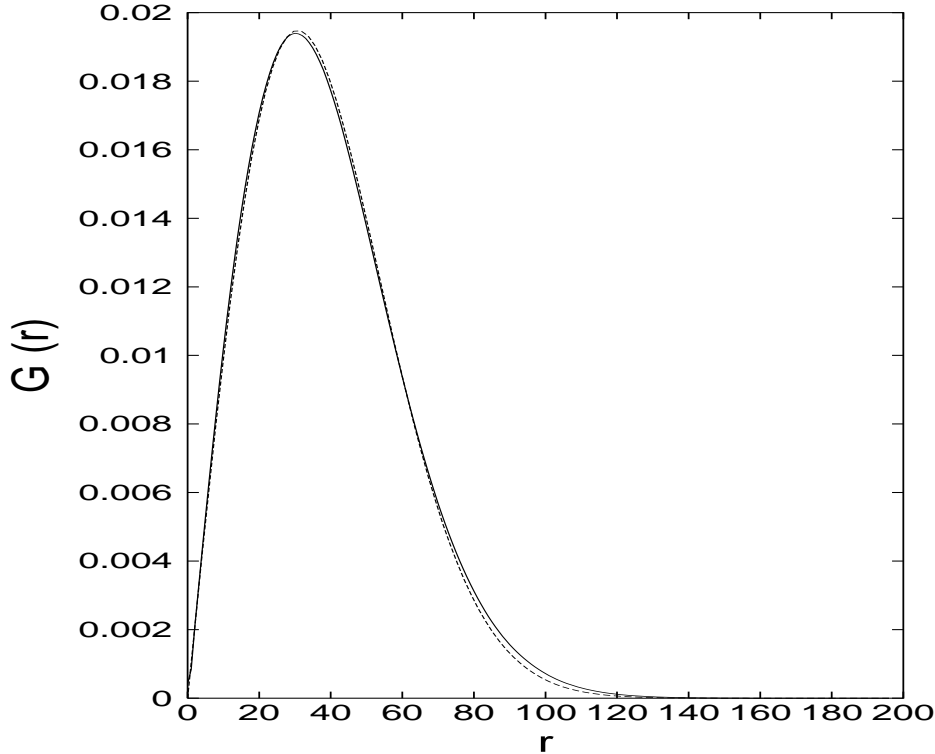,height=10cm,width=12cm}
\end{center}
\caption{\label{f4}The correlation function $G(r)$ for the minbu tree in
the elongated phase (solid line), and the best fit of the form
$G(r) = a r e^{-c r^2 / N}$ (dotted line).}
\end{figure}
\begin{figure}[t]
\begin{center}
\epsfig{file=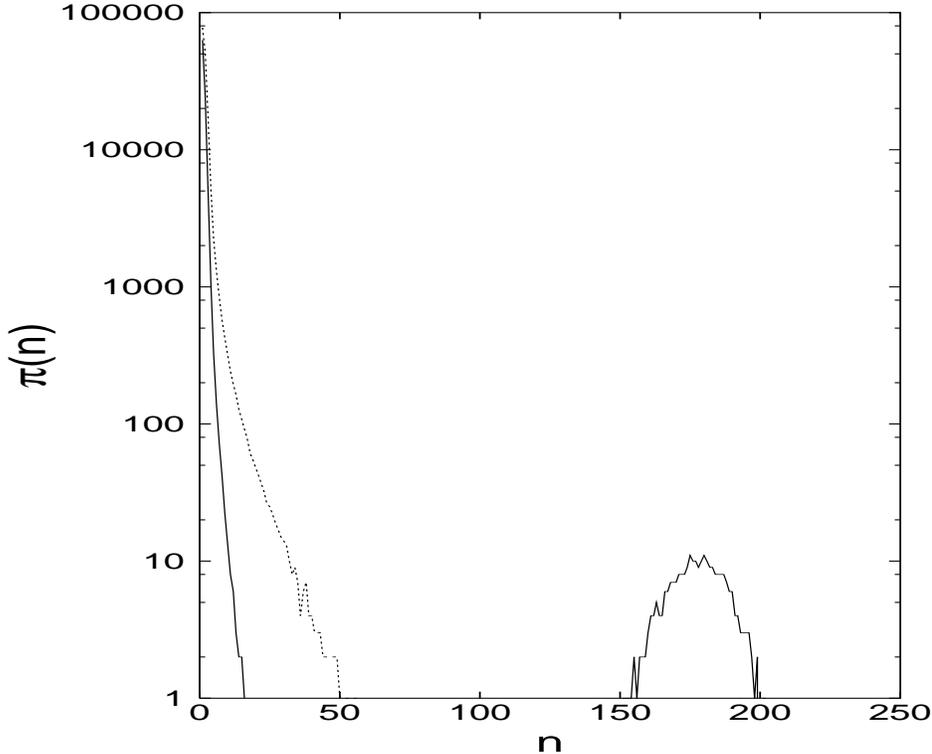,height=10cm,width=12cm}
\end{center}
\caption{\label{f5}The distribution $\pi(n)$ of branching numbers $n$  
on the minbu tree for $\kappa_2 = 1.0$ 
(solid line) and $\kappa_2 = 1.4$ (dotted line).}
\end{figure}
\begin{figure}[t]
\begin{center}
\epsfig{file=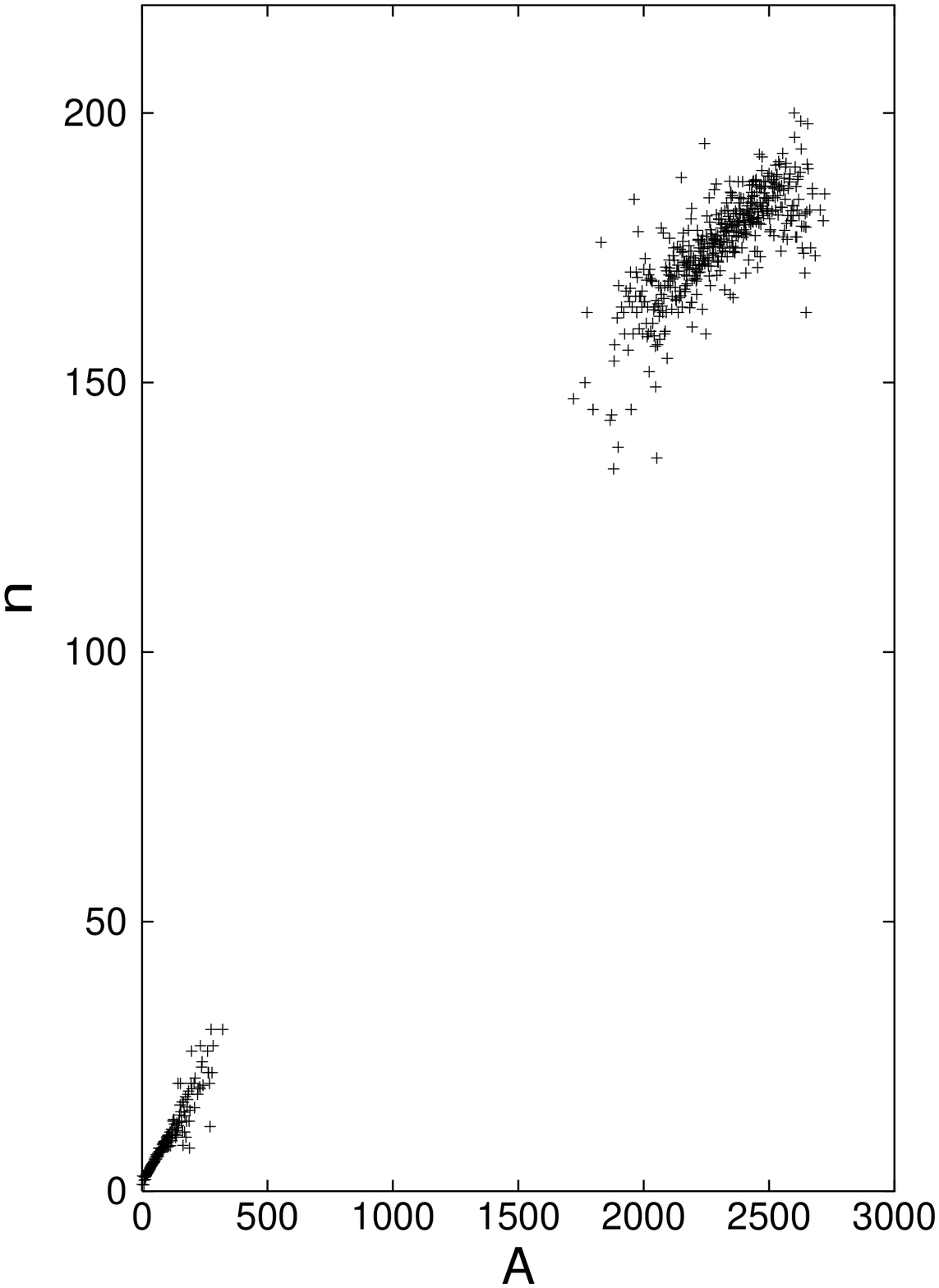,height=10cm,width=12cm}
\end{center}
\caption{\label{f6} Minbu branching numbers $n$ as a function of the minbu size $A$
in the crumpled phase ($N_4 = 4000$, $\kappa_2 = 1.0$).}
\end{figure}
For the ensemble of minbu trees one can define geometrical quantities
analogous to those in simplical gravity. The geodesic distance between
two vertices on the tree is just the number of links between them. (On
the tree there is only one path leading from any one point to any
other). The local quantity which plays the role of the curvature is
the vertex order. One expects the vertex orders ({\em i. e.}, the numbers of 
neighbouring minbus for each baby universe) to be almost independent
from each other, and the main effect of their interaction
to come from the constraint. To check this we measured the
vertex--vertex correlation function on the minbu trees in the elongated phase
($\kappa_2 > \kappa_2^c$) and compared it to the formula for
the correlator for branched polymers~: $ G(r) \sim r e^{-cr^2/N}$ (\cite{aj2}),
where $r$ is the geodesic distance as defined above, and $N$ is the number
of vertices. The results are presented in figure \ref{f4} and 
one sees that the data fits well the function $G(r)$. Moreover, measuring 
the size distribution of minbus one can
directly find the entropy exponent $\gamma = 1/2$. This means that the
ensemble of minbu trees is in a genuine branched polymer phase,
and that the constrained mean--field theory works in the whole
region of $\kappa_2$ above the critical point. It
is now very natural to expect that it will work at the critical point
as well; and, as can be seen in figure \ref{f5} which shows the
distribution of vertex orders for different values of $\kappa_2$,
indeed it does. In the elongated phase, the curve falls off very
rapidly for higher branching numbers.  In the crumpled phase, we see
the same separated peak as we did in the branched polymer model, where
it corresponds to the bush root. This singular root appears exactly in
the critical region, and we found its order to grow linearly
with $N$, as expected. The trees change to bushes at this
point, and we see that the phase transition is indeed driven by the
constrained mean--field mechanism.

Finally, we will show that this transition to the crumpled phase
coincides with the appearance of a mother universe, by which we
mean a part of the universe that has a much larger area than
all other minbus. It is very intuitive to argue that the minbu
which corresponds to the root of the underlying tree
structure must have a relatively large size to allow for many
outgrowths on it. Indeed, figure \ref{f6} shows the clear correlations
between the size of a minbu and the order of the
corresponding vertex. By the size of the minbu we here mean the size of
that part of the minbu obtained by cutting out all direct outgrowths
and then 'sewing up' each of the resulting holes with a
four--simplex. We also find that the mother universe is not only the
largest minbu on the tree, but also contains the singular vertex. In
fact, this vertex is the reason why the mother universe has such a
large volume, simply because the neighbourhood of the singular vertex
contains so many four--simplices. This correlation between the
appearance of singular vertices and the change in the minbu structure
has already been observed in \cite{bbkp2}.

\section*{Simplicial gravity}

Comparison of figures \ref{f1},\ref{f2},\ref{f3} with figure \ref{f4} and
those in \cite{hin} clearly shows a great similarity between
simplicial 4d gravity, minbu trees, and branched polymers, not only in the
elongated phase, but also beyond the transition into the crumpled phase.
It seems apparent that in all these cases the constrained mean--field
mechanism plays the crucial role in the transition.
In fact, branched polymers are by construction the realization
of the mean--field theory (\ref{znm}). The situation is slightly different
for the ensemble of minbu trees, which is an effective theory; here
one has to argue that the formula (\ref{znm}) provides a good
approximation, and that the corrections to this formula coming from
the two-- (or multi--) vertex action play a secondary role. 
However, it should indeed be very plausible that the number of outgrowths
from two different minbus are weakly correlated and not in a position 
to change the main mean--field effect. Very similarly, if one uses
the concept of an effective one--vertex action for the orders of vertices
in simplicial gravity, one ends up with the constrained mean--field theory
as well, due to the constraint on the total sum of vertex orders in the
canonical ensemble. In the next two sections, we will discuss numerical
results from simulations of 2d and 4d simplicial gravity to support our
argument.  Finally, as an application of the constrained mean--field
scenario, we will use it to explain why singular links are found in
the crumpled phase of 4d gravity but singular triangles are not.

\section*{2d}

In 2d simplicial gravity, the model has been studied with $d$ 
gaussian fields coupled to the triangulation, with a
measure term that directly depends on the vertex order
as $n^{\beta}$ \cite{k}. Simulations show that by changing 
$\beta$ or $d$ one can drive the system from the elongated to 
the crumpled phase (called collapsed phase in \cite{k}).
A typical configuration in this phase looks just like a
two--layer pancake. It consists of two discs, each of which is concentrated
around a singular vertex of very high order and glued to the other one
along its perimeter. In the ideal case, when there is no roughness on the
discs, each of them is a set of $n$ triangles, all of which have a common
point of order $n$ in the disc's center. After gluing the discs together,
the $n$ points on the perimeter will be of order $4$. One recognizes the
typical constrained mean--field situation~: a constraint on the sum of vertex
orders -- $\sum_i o_i = 3N$, where $N$ is the number of triangles -- and
an effective action, obtained by integrating out the gaussian sector and
adding the one--vertex term $\beta \log n$ from the measure
force the system to produce singular vertices. When the system grows in size,
the orders of the singular points do likewise. Unlike the branched polymer
situation, however, the geometrical structure of the two--dimensional manifold
forces the appearance of two singular vertices via local geometrical
correlations. In fact, it was shown in \cite{adfo} that by putting a
large number of gaussian fields on the vertices of the triangulation, 
one effectively introduces the measure $n^{-d/2}$.

\section*{4d}

In four dimensions the main effect will be that of changing the
occupation density. As mentioned before, this can trigger the phase
transition as well (see figure \ref{f3}). By changing the coupling 
$\kappa_2$, one changes the average vertex order. 
This value is equal to the average
number of four--simplices to which any vertex belongs and can be
calculated as $\langle o \rangle = 5N_4/N_0$, where $N_4$ is the
number of four--simplices and $N_0$ denotes the number of points.
(The factor $5$ comes from the fact that each four--simplex has five
vertices~: $\sum_i o_i = 5 N_4$).
By decreasing $\kappa_2$ while keeping $N_4$ constant
({\em i. e.}, while staying in the canonical ensemble), one lowers $N_0$. In
the language of the $N$--boxes, $M$--balls model, we increase the
density $\rho=M/N$ by changing the number of boxes. For large
$\kappa_2$ in the crumpled phase, $N_0 \approx N_4/4$, and the
average vertex order becomes $\langle o \rangle \approx 20$.
For $\kappa_2 = 0$ in the crumpled phase, for
$N_4 =32000$ for example, $N_0$ is of order $1000$, and hence the
average order is $\langle o \rangle \approx 160$, which favours a phase
with singular vertices. In fact, for $N_4=32000$ the transition
already takes place for $\kappa_2 \sim 1.258$, where the average order
is $\langle o \rangle \approx 30$; therefore, at $\kappa_2=0$ one
should expect a very strong domination of singular vertices. Indeed,
this turns out to be true. Note that for fixed $N_4$ and $\kappa_2$
the number $N_0$ is not fixed, so strictly speaking this is not the
balls--in--boxes model we have been considering so far.  Therefore one
should in principle consider $N_0$ as a random variable with a certain
smeared distribution. However, the width of this distribution is
relatively small, and one expects that its effects can be neglected here.

In terms of the constrained mean--field mechanism, we can now also try
to explain why singular links appear in the crumpled phase of 4d simplicial
gravity, but singular triangles do not. As we already know, in the crumpled
phase the system favours configurations with as few singular vertices as
possible, because these can then have the greatest possible order. Consider
the neigbourhood of a singular vertex. It has the topology of a 4d ball
and a three--dimensional boundary with the topology of a 3d sphere.
The number of tetrahedra on the boundary corresponds to the order
of the singular vertex in the centre of the ball, and the number
of points on the boundary corresponds to the number of links emerging 
from the sphere. The boundary has its own curvature, since on a
three--dimensional sphere one can keep the number of tetrahedra fixed
and still lower the number of points. This is exactly what happens when
one decreases $\kappa_2$, because this lowers the number of points on the
4d manifold in general, and thus also in the neigbourhood of the singular
vertex. When the number of points on the boundary of this neighbourhood
becomes small enough, the density--forced transition of the constrained
mean--field mechanism is triggered, and another singular vertex
appears. Since this point being part of the neighbourhood of the first
singular vertex means that there will be a link connecting the two, one has
just produced a singular link.

By the same argument, however, one cannot create a
singular triangle in this manner~: the boundary of a link's
neighbourhood has the topology of a 2d sphere, for which the ratio
of triangles and points is fixed by the Euler relation. But if this ratio
cannot be changed, it is obviously impossible to trigger a density--forced
transition, and there is thus no reason why one of the points on the
boundary of the neigbourhood of a singular link should become another
singular vertex. Of course, that still leaves us with the possibility
that a singular triangle might appear directly, without referring to
a singular link already existing on the manifold.
But this turns out to be equally unlikely, mainly
because the average triangle order is a rather slow--changing function of
$\kappa_2$. More specifically, for the average triangle order we have
$\langle o \rangle = 10 N_4/N_2$, where $N_2$ is the number of triangles
on a configuration. Numerical simulations show that for large $N_4$ the
ratio $N_2/N_4$ is asymptotically limited by an upper bound of $2.5$ in
the elongated phase and a lower bound of $2.0$ in the crumpled phase; thus,
over the whole range of $\kappa_2$, the average triangle order varies only
in the very narrow range $\langle o \rangle \in (4,5)$. Additionally, this
range is very close to the minimal triangle order, which is $3$. Compare
this with the situation for vertices, which have a minimal order of $5$
but an average order growing from $20$ up to extremely large numbers,
and it becomes clear why singular triangles should be expected to appear
much more reluctantly than singular vertices or links.

Note that following the scenario outlined here, there might -- or even
should -- be a range of $\kappa_2$ values below the critical point where
singular vertices are already present but singular links are yet missing
because the total number of points is still too large. It would certainly
be worthwhile to check this.

\section*{Summary}

To summarize, in this paper we proposed the constrained mean--field scenario
as a very simple explanation of the transition mechanism between the crumpled
and elongated phases in models of random geometries. Put shortly, one might
say that the elongated phase is entropy dominated while the crumpled phase
is, in contrast, energy dominated. In the crumpled phase, singular spikes
appear in the dressed one--point distributions; depending on context these
correspond to either the singular vertices, the mother universe, or the root
of the bush. This phase results only from the constraint inherent in a
canonical ensemble and is therefore to some extent pathological. For
instance, it is the fast increase of the singularity with the system volume,
caused only by the constraint, that is responsible for the effectively
infinite Hausdorff dimension of geometries in this phase. Since the ensemble
of simplicial gravity that is (hopefully) related to the continuum physics
is the grand canonical one, this would imply that the crumpled phase has
no physical meaning. Since, however, one cannot directly study the
unconstrained ensemble on a computer, one should instead try to keep the
system in its natural phase, namely the entropy--dominated one. One way
of doing this would be to suppress the singular vertices by introducing a
cut--off for the maximal vertex order in the distribution. Alternatively,
as has already been done in 2d simulations, one could modify the integration
measure by adding terms $n^{\beta}$ (where $n$ is the vertex order),
and keep looking for some kind of non--trivial physics in the elongated phase.
A third possibility would be to add fermionic degrees of freedom to prevent
the vertex--order condensation of the system.
Note that the transition to the crumpled phase 
is rather pathological and (as in 2d) should not be of any 
physical importance in the context of continuum gravity. 
In any case, it is known to be highly 
non--universal~: as pointed out in \cite{bb} the order of 
the transition can be easily altered by changing 
the one--vertex action.

The authors are grateful to Sven Bilke for many valuable discussions.
We are indebted to the IC3A and HLRZ J\"{u}lich for the computer time.
P.B. thanks the Stichting voor Fundamenteel Onderzoek
der Materie (FOM) for financial support. Z.B. has
benefited from the financial support of the
Deutsche Forschungsgemeinschaft under the contract
Pe 340/3--3. The work was partially supported by the KBN grant 2P03B 196 02.

\end{document}